\documentstyle[amsmath,amssymb,amscd,aps,pra]{revtex}

\newtheorem{defi}{Definition}
\newtheorem{lemma}[defi]{Lemma}
\newtheorem{satz}[defi]{Theorem}
\newtheorem{cor}[defi]{Corollary}
\newtheorem{bem}[defi]{Note}

\newtheorem{expl}[defi]{Example}

\newtheorem{conj}[defi]{Conjecture}

\newcommand{\proof}{{\emph{Proof.\ }}}
\newcommand{\qed}{\hfill $\Box$}
\newcommand{\tr}{{\operatorname{Tr}\,}}
\newcommand{\id}{{\operatorname{id}}}
\newcommand{\supp}{{\operatorname{supp}\,}}
\newcommand{\bra}[1]{{\langle{#1}|}}
\newcommand{\ket}[1]{{|{#1}\rangle}}
\newcommand{\C}{{\mathbb{C}}}

\newcommand{\alg}[1]{{\mathfrak{#1}}}
\newcommand{\fset}[1]{{\mathcal{#1}}}

\begin{document}


\title{Languages of Quantum Information Theory}
\author{Andreas Winter\thanks{Electronic address: \texttt{winter@mathematik.uni-bielefeld.de}}}
\address{SFB 343, Fakult\"at f\"ur Mathematik\\ Universit\"at Bielefeld, Postfach 100131, 33501 Bielefeld}
\date{July 31, 1998}

\maketitle


\begin{abstract}
  This note will introduce some notation and definitions for information theoretic
  quantities in the context of quantum systems, such as (conditional) entropy and
  (conditional) mutual information. We will employ the natural C${}^*$--algebra
  formalism, and it turns out that one has an allover dualism of language: we can
  define everything for (compatible) observables, but also for (compatible)
  C${}^*$--subalgebras. The two approaches are unified in the formalism of
  quantum operations, and they are connected by a very satisfying
  inequality, generalizing the well known Holevo bound.
  Then we turn to communication via (discrete memoryless) quantum channels:
  we formulate the Fano inequality, bound the capacity region
  of quantum multiway channels, and comment on the quantum broadcast channel.
\end{abstract}


\section{Introduction}
  \label{sec:intro}
  After the beginnings of quantum information theory in the
  sixties~\cite{historical:beginnings}, and Holevo's now
  widely known investigations of the
  seventies~\cite{holevo:bound,holevo:channels,holevo:problems}
  today there is again a tremendous
  interest in this field. This interest focuses on two areas which may be described,
  sightly abusing language introduced by Holevo twenty years ago~\cite{holevo:channels},
  as \emph{classical--quantum}
  problems on the one hand, and \emph{quantum--quantum} problems on the other, and
  it mostly derives from the latter, as these include all problems of (quantum)
  information processing inside a quantum computer or memory. Whereas this area
  (which is charcterized by its attention to \emph{entanglement})
  poses many new and beautiful, and also very difficult problems, the present note
  is concerned wholly with the former area (though it is by now not altogether
  clear how to separate these two worlds, cf. e.g. opinion uttered by
  Adami and Cerf~\cite{adami:cerf:kaballa}). We take the view that classical--quantum
  problems are those in which classical information has to be stored in or sent trough
  some quantum system. Examples from recent work are the determination of the quantum
  channel capacity for fixed input
  states~\cite{hausladen:et:al,holevo:capacity,schumacher:westmoreland},
  quantum cryptographic protocols~\cite{bb:84,mayers:yao}, and
  entanglement enhanced transmission (superdense coding)~\cite{bennett:wiesner}.\\
  Our approach is somewhat reminiscent of ``quantum probability'' through its formulation
  in terms of C${}^*$--algebras and its emphasis on observable operators (which
  reflects our dwelling in the classical--quantum area), but we cannot respect the
  bounds of this field: we will use positive operator valued measures (instead
  of unbounded selfadjoint operators), and we will
  consider quantum operations, both quite uncommon in noncommutative probability.
  Finally it should be noted that we hardly present any new concepts or results
  --- our contribution lies in introducing a reasonable and efficient
  calculus.\par
  The outline of the paper is as follows: in section~\ref{sec:povm} we will
  basically recall the language of C${}^*$--algebras, completely positive maps,
  positive operator valued measures, and the notion of compatibility. In the
  following sections~\ref{sec:observable} and \ref{sec:subalgebra} we will
  define various information theoretic quantities, first for observables, second
  for $*$--subalgebras. In section~\ref{sec:common} we will unify these approaches
  using completely positive C${}^*$--algebra maps, and can give meaning to some hybrid
  expressions in section~\ref{sec:pidgin}. The observable and subalgebra
  notions will be brought together in section~\ref{sec:bound} where we prove
  an information inequality in generalization of the Holevo bound.
  Up to this point the work consists in the definition of concepts
  and information theoretic quantities, and proving some simple
  numerical relations. The last section~\ref{sec:qmwc} will discuss the
  application of these concepts to quantum channels, stating a Fano inequality,
  and determining a bound on the capacity region of the quantum multiway channel.
  We conclude by making some observations for the quantum broadcast channel.\par
  About notation: finite sets will be denoted $\fset{A},\fset{B},\ldots$,
  the functions $\exp$ and $\log$ are always to basis $2$.

\section{Mathematical description of quantum systems}
  \label{sec:povm}
  \setcounter{defi}{0}
  In classical probability theory one has generally two ways of seeing things: either
  through distributions (and the relation of their images, mostly marginals), or through
  random variables (with a common distribution). Both ways have their merits (though
  random variables are considered more elegant), but basically they are equivalent,
  in particular none lacks anything without the other. Things are different in quantum
  probability, and we will take the following view: the analog of a distribution is
  a density operator on some complex Hilbert space, whereas the analog
  of random variables are \emph{observables}, defined below. With density operators
  alone we can study physical processes transforming them, but every experiment
  involves some observable. Studying observables one usually fixes the underlying
  density operator (as the statistics of the experiments depend on the latter),
  but this falls short of not appropriately reflecting our manipulating quantum
  states, or having several alternative states.\\
  For the following we refer to textbooks on C${}^*$--algebras
  like Arveson~\cite{arveson:invitation}, Dixmier~\cite{dixmier:C:algebra}, and standard
  references on basic mathematics of quantum mechanics: Davies~\cite{davies:opensystems},
  Kraus~\cite{kraus:states:etc}, and the more advanced~\cite{holevo:quantum:statistics}
  by Holevo.

  \subsection{Systems and their states}
    A C${}^*$--\emph{algebra} with unit is a Banach space $\alg{A}$ which is also a
    $\C$--algebra with unit $\openone$, and a ${\mathbb{C}}$--antilinear involution $*$, such that
    $$\|AB\|\leq \|A\|\|B\|,\qquad \|A^*\|^2=\|A\|^2=\|AA^*\|$$
    These algebras will be the mathematical models for quantum systems, and subsystems
    are simply $*$--subalgebras.\\
    The set $\alg{A}^+$ of $A\in\alg{A}$ that can be written as $A=BB^*$ is called the
    \emph{positive cone} of $\alg{A}$ which is norm closed, and induces a partial
    order $\leq$. By the famous Gelfand--Naimark--Segal representation
    theorem (see e.g.~\cite{arveson:invitation}) every C${}^*$--algebra
    is isomorphic to a closed $*$--subalgebra of some
    $\alg{L}({\cal H})$, the algebra of bounded linear operators on the Hilbert space ${\cal H}$.
    In this note all C${}^*$--algebras will be of finite dimension. It is known that those
    algebras are isomorphic to a direct sum of $\alg{L}({\cal H}_i)$
    (see e.g. Arveson~\cite{arveson:invitation}).\footnote{
      It is certainly the case that most of the material presented may be generalized
      to infinite dimensional algebras (see e.g. Ohya/Petz~\cite{ohya:petz}). We decided not to try
      for several reasons: one is that in information theory the interesting things already happen
      in the discrete and even finite domain, another (decisive) that the present author is
      only a stumbling beginner in the vast field of C${}^*$--algebras. At least it seems
      clear that the bulk of the things presented here carries over to algebras which
      are isomorphic to countable sums of full (bounded) operator algebras of separable Hilbert
      spaces: there we have trace, well behaved tensor products, and the Schatten decomposition
      (diagonalization) of density operators.}
    This includes as extremal cases the algebras $\alg{L}({\cal H})$, and the commutative
    algebras $\C\fset{X}$ over a finite set $\fset{X}$. In particular we have on every such
    algebra a well defined and unique \emph{trace functional}, denoted $\tr$, that assigns
    trace one to all minimal positive idempotents.\\
    A \emph{state} on a C${}^*$--algebra $\alg{A}$ is a positive $\C$--linear functional
    $\rho$ with $\rho(\openone)=1$. Positivity here means that its values on the positive cone
    are nonnegative. Clearly the states form a convex set $\alg{S}(\alg{A})$ whose extreme points are called
    \emph{pure} states, all others are \emph{mixed}. One can easily see that every state
    $\rho$ can be represented uniquely in the form
    $\rho(X)=\tr(\hat{\rho} X)$ for a positive, selfadjoint element $\hat{\rho}$ of $\alg{A}$
    with trace one (such elements are called \emph{density operators}).
    In general this is only true for so--called \emph{normal} states, which means that
    for an increasing sequence $A_n$ converging in norm to $A$ the values $\rho(A_n)$
    converge to $\rho(A)$.
    In the sequel we will therefore make no distinction between $\rho$ and its
    density operator $\hat{\rho}$.
    The set of operators with finite trace will be denoted $\alg{A}_*$, the
    \emph{trace class} in $\alg{A}$ which contains the states and is a two--sided
    ideal in $\alg{A}$, the {\sc Schatten}--ideal~\cite{schatten:operators}.
    $\tr(\rho A)$ then defines a real bilinear and nondegenerate pairing
    of $\alg{A}_{*s}$ and $\alg{A}_s$, the selfadjoint parts of $\alg{A}_*$ and $\alg{A}$
    which makes $\alg{A}_s$ the dual of $\alg{A}_{*s}$.
    Notice that in this sense pure states are equivalently
    described as minimal selfadjoint idempotents of $\alg{A}$.

  \subsection{Observables}
    Let ${\cal F}$ be a $\sigma$--algebra on some set $\Omega$, $\alg{X}$ a C${}^*$--algebra.
    A map $X:{\cal F}\longrightarrow \alg{X}$ is called a \emph{positive operator valued
    measure} (POVM), or an \emph{observable}, with values in $\alg{X}$ (or on $\alg{X}$), if:
    \begin{enumerate}
      \item $X(\emptyset)=0,\  X(\Omega)=\openone$.
      \item $E\subset F$ implies $X(E)\leq X(F)$.
      \item If $(E_n)_n$ is a countable family of pairwise disjoint sets in ${\cal F}$ then
        $X(\bigcup_n E_n)=\sum_n X(E_n)$ (in general the convergence is to be
        understood in the weak topology: for every state its value at the left
        equals the limit value at the right hand side).
    \end{enumerate}
    If the values of the observable are all projection operators and $\Omega$ is
    the real line one speaks of a
    \emph{spectral measure} or a \emph{von Neumann observable}.\footnote{Strictly
      speaking this term only applies to the expectation of the measure
      (in general an unbounded operator), but this in turn by the spectral theorem
      determines the measure.}
    An observable $X$ together with a state $\rho$ yields a probability measure
    $P^X$ on $\Omega$ via
    $$P^X(E)=\tr(\rho X(E))$$
    In this way we may view $X$ as a random variable with values in $\alg{X}$, its
    distribution we denote $P_X$ (note that $P_X$ may not be isomorphic to $P^X$: if
    $X$ takes the same value on disjoint events, which means that $X$ introduces randomness
    by itself).\\
    Two observables $X$, $Y$ are said to be \emph{compatible}, if
    they have values in the same algebra and
    ${XY}={YX}$ elementwise, i.e. for all $E\in{\cal F}_X$,
    $F\in{\cal F}_Y$: $X(E)Y(F)=Y(F)X(E)$
    (Note that it is possible for an observable not to be compatible with itself).
    By the way, the term \emph{compatible}
    may be defined in obvious manner for arbitrary sets or collections of operators, in
    which meaning we will use it in the sequel.
    If $X,Y$ are compatible we may define their \emph{joint observable}
    ${XY}:{\cal F}_X\times{\cal F}_Y\longrightarrow\alg{X}$
    mapping $E\times F$ to $X(E)Y(F)$ (this defines the product mapping
    uniquely just as in the classical case of product measures).
    In fact we can analogously define the
    joint observable for any collection of pairwise compatible observables.\footnote{Observe
      however that in general a joint observable might exist for non--compatible (i.e.
      non--commuting) observables. The operational meaning of this is that there is a
      common refinement of the involved observables. If they commute then this certainely is
      possible as demonstrated, but commutativity is not necessary.}
    As the random variable of a product ${XY}$ we will take $X\times Y$,
    rather than $XY$ itself, with values in $\alg{X}\times\alg{X}$ (because the same product
    operator may be generated in two different ways which we want to distinguish).
    To indicate this difference we will sometimes write $X\cdot Y$ for the product.\par
    Note that here we can see the reason why we cannot just consider all observables as
    random variables (and forget about the state): they will not have a joint distribution,
    at first of course only by our definition. But Bell's theorem~\cite{bell:theorem}
    shows that one comes into serious trouble if one tries to allow a joint distribution
    for noncompatible observables. Conversely we see why
    we cannot do without observables, even though
    $\rho$ contains all possible information: the crux is that we cannot access it due to
    the forbidden noncompatibel observables (a good account of this aspect of quantum
    theory is in~\cite{peres:quantum}). \par
    From now on all observables will be \emph{countable},
    i.e. w.l.o.g. are they defined on a countable $\Omega$
    with $\sigma$--algebra $2^\Omega$. This
    means that we may view an observable $X$ as a resolution of $\openone$ into a countable
    sum $\openone=\sum_{j\in\Omega} X_j$ of positive operators $X_j$.\\
    If $\alg{A}_1,\alg{A}_2$ are subalgebras of $\alg{A}$, they are compatible if they
    commute elementwise (again note, that a subalgebra need not not be compatible with itself:
    in fact it is iff it is commutative). In this case the closed subalgebra generated
    (in fact: spanned) by the products $A_1A_2$, $A_i\in\alg{A}_i$ is denoted $\alg{A}_1\alg{A}_2$.

  \subsection{Quantum operations}
    Now we describe the transformations between quantum systems: a $\C$--linear
    map $\varphi:\alg{A}_2\rightarrow\alg{A}_1$ is called a \emph{quantum operation}
    if it is completely positive (i.e. positive, so that positive elements have positive
    images, and also the $\varphi\otimes\id_n$ are positive, where $\id_n$ is the identity
    on the algebra of $n\times n$--matrices), and unit preserving. These
    maps are in $1$--$1$ correspondence with their (pre--)adjoints $\varphi_*$
    by the trace form, mapping states to states, and being completely positive and
    trace preserving.\footnote{In general this is only true if we restrict $\varphi$
      to be a \emph{normal} map, see Davies~\cite{davies:opensystems}.}
    Since here we restrict ourselves to finite dimensional algebras
    the adjoint map simply goes from $\alg{A}_1$ to $\alg{A}_2$, but to keep things well
    separated (which they actually are in the infinite case) we write the adjoint
    as $\varphi_*:\alg{A}_{1*}\rightarrow\alg{A}_{2*}$, the dual map (in fact
    we consider this as the primary object and the operator maps as their adjoint, which
    is the reason for writing subscript $*$).
    Notice that $\varphi_*$ is sometimes considered as restricted to
    $\varphi_*:\alg{S}(\alg{A}_1)\rightarrow\alg{S}(\alg{A}_2)$. A
    characterization of quantum operations is by the Stinespring
    dilation theorem~\cite{stinespring:thm}:
    \begin{satz}[Dilation]
      \label{satz:dilation}
      Let $\varphi:\alg{A}\rightarrow\alg{L}({\cal H})$ a linear map of C${}^*$--algebras.
      Then $\varphi$ is completely positive if and only if
      there exist a representation $\alpha:\alg{A}\rightarrow\alg{L}({\cal K})$,
      with Hilbert space ${\cal K}$, and a bounded linear map
      $V:{\cal H}\rightarrow{\cal K}$ such that
      $$\forall A\in\alg{A}\qquad \varphi(A)=V^*\alpha(A)V$$
    \end{satz}
    For proof see e.g.~\cite{davies:opensystems}.

  \subsection{Entropy and divergence}
    We will talk about information theory, so we need a concept of entropy: the
    von Neumann entropy $H(\rho)=-\tr(\rho\log\rho)$ (introduced in~\cite{von:neumann:entropy})
    of a state $\rho$
    (which reduces to the usual Shannon entropy for a commutative algebra because then
    a state is nothing but a probability distribution). For states
    $\rho,\sigma$ also introduce the I--divergence (first defined
    by Umegaki~\cite{umegaki:divergence}), or simply divergence
    as $D(\rho\|\sigma)=\tr(\rho(\log\rho-\log\sigma))$ with the convention that this is $\infty$ if
    $\supp\rho\not\leq\supp\sigma$ ($\supp\rho$ being the \emph{support}
    of $\rho$, the minimal selfadjoint idempotent $p$ with $p\rho p=\rho$).
    For properties of these quantities we will often refer to~\cite{ohya:petz},
    and to~\cite{wehrl:entropy}.
    Two important facts we will use are
    \begin{satz}[Klein inequality]
      \label{satz:klein}
      For positive operators $\rho,\sigma$ (not necessary states)
      $$D(\rho\|\sigma)\geq \frac{1}{2}\tr(\rho-\sigma)^2+\tr(\rho-\sigma)$$
      In particular for states the divergence is nonnegative.
    \end{satz}
    \proof See~\cite{ohya:petz}. \qed
    \begin{satz}[Monotonicity]
      \label{satz:monoton}
      Let $\rho,\sigma$ be states on a C${}^*$--algebra $\alg{A}$, and $\varphi_*$ a trace preserving,
      completely positive linear map from states on $\alg{A}$ to states on $\alg{B}$. Then
      $$D(\varphi_*\rho\|\varphi_*\sigma)\leq D(\rho\|\sigma)$$
    \end{satz}
    \proof Uhlmann~\cite{uhlmann:monoton}, the situation we are in was already solved
    by Lindblad~\cite{lindblad:monoton}. For a textbook account see~\cite{ohya:petz}. \qed

\section{Observable language}
  \label{sec:observable}
  \setcounter{defi}{0}
  Fix a state on a C${}^*$--algebra, say $\rho$ on $\alg{A}$ and let $X,Y,Z$ compatible
  observables on $\alg{A}$.\\
  By the previous section~\ref{sec:povm} these are then random variables with
  a joint distribution, and one defines entropy $H(X)$, conditional entropy
  $H(X|Y)$, mutual information $I(X\wedge Y)$, and conditional
  mutual information $I(X\wedge Y|Z)$ for these observables as the respective
  quantities for them interpreted as random variables.
  Note however that these depend on the underlying state $\rho$. In case of
  need we will thus add the state as an index, like $H_\rho(X)=H(X)$, etc.\\
  As things are there is not much to say about that part of the theory.
  We only note some useful formulas:
  $$H(X|Y)=\sum_j {\tr(\rho Y_j)H_{\rho_j}(X)},\quad\text{with }
                         \rho_j=\frac{1}{\tr(\rho Y_j)}\sqrt{Y_j}\rho\sqrt{Y_j}$$
  (which is an easy calculation using the compatibility of $X$ and $Y$), and
  \begin{equation*}\begin{split}
    {I}(X\wedge Y) &= H(X)+H(Y)-H({XY})\\
                   &= D(P^{XY}\|P^X\otimes P^Y)=D(P_{X\cdot Y}\|P_X\otimes P_Y)
  \end{split}\end{equation*}
  (which is known from classical information theory).

\section{Subalgebra language}
  \label{sec:subalgebra}
  \setcounter{defi}{0}
  Let $\alg{X},\alg{X}_1,\alg{X}_2,\alg{Y}$ compatible
  $*$--subalgebras of the C${}^*$--algebra
  $\alg{A}$, and $\rho$ a fixed state on $\alg{A}$.\\
  First consider the inclusion map $\imath:\alg{X}\hookrightarrow\alg{A}$ (which is certainly
  completely positive) and its adjoint
  $\imath_*:\alg{A}_*\rightarrow\alg{X}_*$. Define
  $$H(\alg{X})=H_\rho(\alg{X}):=H(\imath_*\rho)$$
  (where at the right hand appears the von Neumann entropy). For example for $\alg{X}=\alg{A}$
  we obtain just the von Neumann entropy of $\rho$. For the trivial subalgebra $\C=\C\openone$
  (which commutes obviously with every subalgebra) we obtain, as expected,
  $H(\C)=0$. The general philosophy behind this
  definition is that $H(\alg{X})$ is the (von Neumann) entropy of the global state
  \emph{viewed through (or restricted to) the subsystem} $\alg{X}$. To reflect this in the
  notation we define $\rho|_{\alg{X}}=\imath_*\rho$.\\
  Now conditional entropy, mutual information, and conditional mutual information
  are defined by reducing them to entropy quantities:
  $$H(\alg{X}|\alg{Y})=H(\alg{XY})-H(\alg{Y})$$
  $$I(\alg{X}_1\wedge\alg{X}_2)=H(\alg{X}_1)+H(\alg{X}_2)-H(\alg{X}_1\alg{X}_2)$$
  \begin{equation*}\begin{split}
    I(\alg{X}_1\wedge\alg{X}_2|\alg{Y}) &=H(\alg{X}_1|\alg{Y})+H(\alg{X}_2|\alg{Y})
                                              -H(\alg{X}_1\alg{X}_2|\alg{Y}) \\
                                        &=H(\alg{X}_1\alg{Y})+H(\alg{X}_2\alg{Y})
                                              -H(\alg{X}_1\alg{X}_2\alg{Y})-H(\alg{Y})
  \end{split}\end{equation*}
  It is not at all clear a priori that these definitions are all well behaved: while
  it is obvious from the definition that the entropy is always nonnegative, this
  is not true for the conditional entropy (as was observed by several authors before):
  if $\alg{A}=\alg{X}\otimes\alg{Y}$ and $\rho$ is a pure entangled state then
  $H(\alg{X}|\alg{Y})=-H(\alg{Y})<0$. This might raise pessimism whether the other two quantities
  also are (at least sometimes) pathological. This they are not, as will be shown in a
  moment:\\
  We have the following commutative
  diagram of inclusions, and the natural multiplication map $\mu$ (which is in fact a $*$--algebra
  homomorphism, and thus completely positive!):
  \begin{equation*}\begin{CD}
    \alg{X}_1                 @=           \alg{X}_1       @=              \alg{X}_1\\
    @VV{\varphi_1}V                      @VV{\imath_1}V                 @VV{\jmath_1}V\\
    \alg{X}_1\otimes\alg{X}_2 @>{\mu}>> \alg{X}_1\alg{X}_2 @>{\jmath}>>     \alg{A}\\
    @AA{\varphi_2}A                      @AA{\imath_2}A                 @AA{\jmath_2}A\\
    \alg{X}_2                 @=           \alg{X}_2       @=              \alg{X}_2
  \end{CD}\end{equation*}
  And hence the corresponding commutative diagram of adjoint maps
  (note that $\varphi_{1*}$ and $\varphi_{2*}$ are just
  partial traces). With this we find
  \begin{equation*}\begin{split}
    {I}(\alg{X}_1\wedge\alg{X}_2) &=H(\alg{X}_1)+H(\alg{X}_2)-H(\alg{X}_1\alg{X}_2)\\
                                  &=H(\jmath_{1*}\rho)+H(\jmath_{2*}\rho)-H(\jmath_*\rho)\\
                                  &=H(\varphi_{1*}\mu_*\jmath_*\rho)+H(\varphi_{2*}\mu_*\jmath_*\rho)
                                                            -H(\mu_*\jmath_*\rho) \\
                             &=D(\mu_*\jmath_*\rho\|\varphi_{1*}\mu_*\jmath_*\rho\otimes\varphi_{2*}\mu_*\jmath_*\rho)
  \end{split}\end{equation*}
  by definition, then by commutativity of the diagram and the fact that $\mu_*$ preserves
  eigenvalues of density operators (because $\mu$ is a surjective $*$--homomorphism, see
  lemma~\ref{lemma:onto:hom} below), the last by direct calculation
  on the tensor product (just as for the classical formula).
  From the last line we see that the mutual information is nonnegative because
  the divergence is, by theorem~\ref{sec:povm}.\ref{satz:klein}
  (we could also have seen this already from the definition by applying
  subadditivity of von Neumann entropy to the second last line, see
  theorem~\ref{sec:bound}.\ref{satz:strong:subadd}).
  \begin{lemma}
    \label{lemma:onto:hom}
    Let $\mu:\alg{A}\rightarrow\alg{B}$ a surjective $*$--algebra homomorphism. Then
    \begin{enumerate}
      \item For all pure states $p\in\alg{S}(\alg{A})$: $\mu(p)$ pure or $0$.
      \item For all $A\in\alg{A}$, $A\geq 0$: $\tr A\geq\tr\mu(A)$.
      \item For pure $p\in\alg{S}(\alg{A})$, $q\in\alg{S}(\alg{B})$:
        $$\mu_*(\mu(p))=p\text{ or }\mu(p)=0,\ \mu(\mu_*(\mu(p)))=\mu(p), \ \mu(\mu_*(q))=q$$
      \item For $\rho\in\alg{S}(\alg{B})$, $\mu_*(\rho)=\sum_i \alpha_i p_i$ diagonalization with
        the $\alpha_i>0$, then $\rho=\sum_i \alpha_i\mu(p_i)$ is a diagonalization.
      \item Conversely every diagonalization of a state on $\alg{B}$ is by $\mu_*$ translated
        into a diagonalization of its $\mu_*$--image.
    \end{enumerate}
  \end{lemma}
  \proof
    \begin{enumerate}
      \item We have only to show that $\mu(p)$ is minimal if it is not $0$: let $q'$ any pure state
        with $q'\leq\mu(p)$. Then
        $$1=\tr(q'\mu(p))=\tr(\mu_*(q')p)\leq\tr(p)=1$$
        So we must have equality which implies $p\leq\mu_*(q')$, but both operators
        are states, so $p=\mu_*(q')$. Because $\mu_*$ is injective this means that there
        is only one pure state $q'\leq\mu(p)$, i.e. $\mu(p)$ is pure.
      \item We may write $A=\sum_i a_ip_i$ with pure states $p_i$ and $a_i\geq 0$.
        Then $\mu(A)=\sum_i a_i\mu(p_i)$ and since pure states have trace $1$ the assertion follows
        from (1).
      \item Let $A\in\alg{A}$, $A\geq 0$. Then
        \begin{equation*}\begin{split}
          \tr(\mu_*(\mu(p))A) &=\tr(\mu(p)\mu(A))=\tr(\mu(p)\mu(A)\mu(p)) \\
                              &=\tr(\mu(pAp))\leq\tr(pAp)=\tr(pA)
        \end{split}\end{equation*}
        Thus $\mu_*(\mu(p))\leq p$. If $\mu(p)\neq 0$ it is a pure state, hence $\mu_*(\mu(p))$ a state which
        forces $\mu_*(\mu(p))=p$. This proves the left formula, the middle follows immediately, and
        for the right observe that we may choose a pure pre--image $p$ of $q$ (in fact that will
        be $\mu_*(q)$, as one can see from (4)).
      \item $\sum_i \alpha_i\mu(p_i)$ is certainly the diagonalization of some positive operator since
        the $\mu(p_i)$ which are not $0$ are by the homomorphism property and by (1) pairwise orthogonal
        pure states. Now observe $\mu(\mu_*(\rho))=\sum_i \alpha_i\mu(p_i)$ and
        $$\mu_*(\rho)=\mu_*(\mu(\mu_*(\rho)))=\sum_i \alpha_i\mu_*(\mu(p_i))\leq\sum_i \alpha_i p_i=\mu_*(\rho)$$
        hence equality, i.e. all $\mu(p_i)$ are pure. From
        $$\mu_*(\rho)=\sum_i \alpha_i\mu_*(\mu(p_i))=\mu_*(\sum_i \alpha_i\mu(p_i))$$
        and injectivity of $\mu_*$ the assertion follows.
      \item This is a direct consequence of (3) and (4).
        \qed
    \end{enumerate}
  \medskip\noindent
  For the conditional mutual information we have to do somewhat more (yet from the definition
  we see that its positivity will have something to do with the strong subadditivity
  of von Neumann entropy, see theorem~\ref{sec:bound}.\ref{satz:strong:subadd}): \\
  Consider the following commuative diagram:
  \begin{equation*}\begin{CD}
    \alg{Y} @>{\varphi_1}>> \alg{X}_1\otimes\alg{Y}               @>{\mu_1}>> \alg{X}_1\alg{Y}\\
      @|                       @VV{\varphi_1'}V                                 @VV{\jmath_1}V\\
    \alg{Y} @>{\varphi}>> \alg{X}_1\otimes\alg{X}_2\otimes\alg{Y} @>{\mu}>>   \alg{X}_1\alg{X}_2\alg{Y}
                                                                                      @>{\jmath}>> \alg{A}\\
      @|                       @AA{\varphi_2'}A                                 @AA{\jmath_2}A\\
    \alg{Y} @>{\varphi_2}>> \alg{X}_2\otimes\alg{Y}               @>{\mu_2}>> \alg{X}_2\alg{Y}
  \end{CD}\end{equation*}
  All maps there are completely positive, $\mu,\mu_1,\mu_2$ being $*$--homomorphisms.
  Thus the adjoints of the various $\varphi$'s are partial traces
  and with $\sigma=\mu_*\jmath_*\rho$:
  $H(\alg{X}_1\alg{X}_2\alg{Y})=H(\sigma)$, $H(\alg{X}_1\alg{Y})=H(\tr_{\alg{X}_2}\sigma)$,
  $H(\alg{X}_2\alg{Y})=H(\tr_{\alg{X}_1}\sigma)$,
  $H(\alg{Y})=H(\tr_{\alg{X}_1\otimes\alg{X}_2}\sigma)$
  (where we have made use of lemma~\ref{lemma:onto:hom} several times),
  and we can indeed apply strong subadditivity.\\
  Finally let us remark the nice formulas
  $$H(\alg{X})=H(\alg{X}|\C),\qquad I(\alg{X}_1\wedge\alg{X}_2)=I(\alg{X}_1\wedge\alg{X}_2|\C)$$
  \begin{expl}
    \label{expl:tensorproduct}
    \normalfont{
    A very important special case of the definitions of this and the preceding section occurs
    for tensor products of Hilbert spaces $\alg{L}({\cal H}_1\otimes{\cal H}_2)
    =\alg{L}({\cal H}_1)\otimes\alg{L}({\cal H}_2)$, or more generally tensor products
    of C${}^*$--algebras: $\alg{A}=\alg{A}_1\otimes\alg{A}_2$.
    $\alg{A}_1,\alg{A}_2$ are $*$--subalgebras of $\alg{A}$ in the natural way,
    and are obviously compatible. The same then holds for observables $A_i\subset\alg{A}_i$,
    and similarly for more than 2 factors. In this case the restriction $\rho|_{\alg{A}_i}$
    is just a partial trace.
    }
  \end{expl}

\section{Common tongue}
  \label{sec:common}
  \setcounter{defi}{0}
  The languages of the two preceding sections may be phrased in a unified formalism
  (the ``common tongue'') using completely positive C${}^*$--algebra maps (in particular
  those from or to commutative algebras, inclusion maps, and $*$--algebra homomorphisms,
  cf. Stinespring~\cite{stinespring:thm}).\\
  That this is promising one can see from the observation that observables can be interpreted
  in a natural way as C${}^*$--algebra maps: $X:\Omega\rightarrow\alg{A}$
  corresponds by linear extension to $X:\alg{B}(\Omega)\rightarrow\alg{A}$, where
  $\alg{B}(\Omega)=\alg{B}(\Omega,{\cal F})$ is the algebra of bounded measurable
  functions on $\Omega$. We follow the convention that in this algebra $j\in\Omega$ shall
  denote the function that is $1$ on $j$ and $0$ elsewhere, so $X(j)=X_j$, and
  obviously $X_*(\rho)$ equals the distribution $P^X$ on $\Omega$ induced by $X$
  with $\rho$.\\
  Let us also introduce some notation for the observable $X$: the \emph{total} observable
  operation $X_{\text{tot}}:\alg{B}(\Omega)\otimes\alg{A}\rightarrow\alg{A}$ mapping
  $j\otimes A\mapsto \sqrt{Y_j}A\sqrt{Y_j}$,
  its \emph{interior} part
  $X_{\text{int}}=X_{\text{tot}}\circ\imath_{\alg{A}}:\alg{A}\rightarrow\alg{A}$ with
  $A\mapsto\sum_j \sqrt{Y_j}A\sqrt{Y_j}$, and its \emph{exterior} part
  $X_{\text{ext}}=X_{\text{tot}}\circ\imath_{\alg{B}(\Omega)}$
  which coincides with $X$.\par
  Consider compatible quantum operations $\varphi:\alg{X}\rightarrow\alg{A}$,
  $\psi:\alg{Y}\rightarrow\alg{A}$, etc.
  ($\varphi,\psi$ are compatible if their images commute elementwise). In this case
  their \emph{product} is $\varphi\psi:\alg{X}\otimes\alg{Y}\rightarrow\alg{A}$
  mapping $X\otimes Y\mapsto\varphi(X)\psi(Y)$:
  \begin{equation*}\begin{CD}
    \alg{X}                 @>{\varphi}>>              \alg{A}\\
    @V{\varphi_1}VV                                       @|\\
    \alg{X}\otimes\alg{Y}   @>{\exists^!\varphi\psi}>> \alg{A}\\
    @A{\varphi_2}AA                                       @|\\
    \alg{Y}                 @>{\psi}>>                 \alg{A}
  \end{CD}\end{equation*}
  Note that this generalizes the product of
  observables, as well as the product map $\mu$ of subalgebras.\\
  Now simply define $H(\varphi)=H(\varphi_*\rho)$, and
  again the conditional entropy and the informations are defined by reduction to
  entropy, e.g. $H(\varphi|\psi)=H(\varphi\psi)-H(\psi)$, or
  $I(\varphi\wedge\psi)=H(\varphi)+H(\psi)-H(\varphi\psi)$.\\
  For the mutual information observe that (see previous diagram):
  \begin{equation*}\begin{split}
    I(\varphi\wedge\psi) &= D((\varphi\psi)_*\rho\|\varphi_*\rho\otimes\psi_*\rho)\\
                         &= D(\sigma\|\tr_{\alg{Y}}\sigma\otimes\tr_{\alg{X}}\sigma)
                                \qquad\text{ with }\sigma=(\varphi\psi)_*\rho
  \end{split}\end{equation*}
  Note the difference to Ohya/Petz~\cite{ohya:petz}: with them the entropy of an operation
  is related to the mutual information of the operation as a channel
  (see section~\ref{sec:qmwc}).
  With us the entropy of an operation is the entropy of a state ``viewed through''
  this operation (as was the idea with the entropy of a subsystem, and obviously also with
  the entropy of an observable).

\section{Pidgin}
  \label{sec:pidgin}
  \setcounter{defi}{0}
  With the insight of the preceding section we may now form hybrid expressions involving
  observables and subalgebras at the same time: let
  $\imath:\alg{X}\hookrightarrow\alg{A}$, $\jmath:\alg{Y}\hookrightarrow\alg{A}$ $*$--subalgebra
  inclusions, and $X,Y$ observables on $\alg{A}$, all four compatible. Then we have
  $$H(\alg{X}|Y)=H(\imath Y)-H(Y)$$
  $${I}(\alg{X}\wedge Y)=H(\imath)+H(Y)-H(\imath Y)$$
  and lots of others. From the previous section we know that the information quantities are nonnegative,
  but also the entropy conditional on an observable, from the formula
  $$H(\alg{X}|Y)=\sum_j \tr(\rho Y_j)H_{\rho_j}(\alg{X}),\quad\text{with }
                         \rho_j=\frac{1}{\tr(\rho Y_j)}\sqrt{Y_j}\rho\sqrt{Y_j}$$
  But also again there are some expressions which seem suspicious, like
  $$H(X|\alg{Y})=H(X\jmath)-H(\alg{Y})$$
  But due to the inequality of theorem~\ref{sec:bound}.\ref{satz:conditional:entropy}
  in fact it behaves nicely.

\section{Inequalities}
  \label{sec:bound}
  \setcounter{defi}{0}

  \subsection{Entropy}
  \begin{satz}
    \label{satz:strong:subadd}
    For compatible $*$--subalgebras $\alg{A}_1,\alg{A}_2,\alg{A}_3$
    one has:
    \begin{enumerate}
      \item Subadditivity: $H(\alg{A}_1\alg{A}_2)\leq H(\alg{A}_1)+H(\alg{A}_2)$.
      \item Strong subadditivity:
        $H(\alg{A}_1\alg{A}_2\alg{A}_3)+H(\alg{A}_2)\leq H(\alg{A}_1\alg{A}_2)+H(\alg{A}_2\alg{A}_3)$.
    \end{enumerate}
  \end{satz}
  \proof
    Subadditivity is a special case of strong subadditivity: $\alg{A}_2=\C$.
    The latter can be reduced to the familiar form (see e.g. Wehrl~\cite{wehrl:entropy})
    by the same type of argument as we used in section~\ref{sec:subalgebra}
    for the nonnegativity of conditional mutual information...
  \qed
  \begin{satz}
    \label{satz:pure:common:state}
    Let $\alg{X},\alg{Y}$ compatible, $\rho|_{\alg{XY}}$ pure. Then $H(\alg{X})=H(\alg{Y})$.
  \end{satz}
  \proof By retracting the state $\rho$ to $\alg{X}\otimes\alg{Y}$ by the multiplication
  map $\mu:\alg{X}\otimes\alg{Y}\rightarrow\alg{XY}$
  (see lemma~\ref{sec:subalgebra}.\ref{lemma:onto:hom}) we may assume that we have a pure
  state $\rho$ on $\alg{X}\otimes\alg{Y}$. Then the assertion of the theorem is
  $H(\tr_{\alg{X}}\rho)=H(\tr_{\alg{Y}}\rho)$ which is well known (proof via the
  polar decomposition of $\rho$...).  \qed
  \par\medskip
  Another kind of inequality may serve as an operational justification of the definition
  of von Neumann entropy. Call a quantum operation $\varphi:\alg{A}_1\rightarrow\alg{A}_2$
  \emph{doubly stochastic} if it preserves the trace, i.e. for all $A\in\alg{A}_1$:
  $\tr\varphi(A)=\tr A$ (see Ohya/Petz~\cite{ohya:petz}). We will consider the less
  restrictive condition $\tr\varphi(A)\leq\tr A$, and
  for an observable $X$ and subalgebra $\alg{X}$
  let us say they are \emph{maximal in} $\alg{A}$ if $X$ and the inclusion map
  have this property (obviously for the subalgebra this implies doubly stochastic).
  Main examples are: an observable whose atoms are minimal in the target
  algebra, i.e. have only trivial decompositions into positive
  operators, and a maximal commutative subalgebra.
  \begin{satz}[Entropy increase]
    \label{satz:entropy:inequ}
    Let $\varphi:\alg{Y}\rightarrow\alg{X}$ with $\tr\varphi(A)=\tr A$, and
    $\psi:\alg{X}\rightarrow\alg{A}$ quantum operations. Then
    $H(\psi\circ\varphi)\geq H(\psi)$. (Notice that in the physical sense
    the operation $\varphi_*$ is applied \emph{after} $\psi_*$).
  \end{satz}
  Before we prove this let us note
  two important case of equality: Let $\rho=\sum_i\lambda_i p_i$ with mutually orthogonal
  pure states $p_i$, $\lambda_i\geq 0$, $\sum_i p_i=\openone$.
  Then equality holds for the subalgebra generated
  by the $p_i$ (in fact for any subalgebra which contains them),
  and for the observable that corresponds to the $p_i$'s
  resolution of $\openone$.\\
  \emph{Proof of theorem~\ref{satz:entropy:inequ}}.
    Let $\sigma=\psi_*\rho$, we have to prove $H(\varphi_*\sigma)\geq H(\sigma)$.
    From the previous discussion we see that we may assume $\alg{Y}$ to be
    commutative, without changing the trace relation. Let $\sigma=\sum_i \alpha_i p_i$
    a diagonalization with pure states $p_i$ on $\alg{X}$, and
    $q_j$ the family of minimal idempotents of $\alg{Y}$ (which by commutativity
    are othogonal). Then we have decompositions
    $\varphi_* p_i=\sum_j \beta_{ij} q_j$, hence
    $$\varphi_*\sigma=\sum_i \alpha_i\varphi_* p_i
                     =\sum_j\left({\sum_i \alpha_i\beta_{ij}}\right)q_j$$
    Now observe that for all $j$
    $$\sum_i \beta_{ij}=\tr(q_j\sum_i \varphi_* p_i)=\tr((\varphi q_j)\sum_i p_i)
                       =\tr(\varphi q_j)\leq\tr(q_j)=1$$
    and the result follows from the formulas
    $H(\sigma)=H(\alpha_i|i)$, $H(\varphi_*\sigma)=H(\sum_i\beta_{ij}\alpha_i|j)$.
  \qed
  \par\medskip
  Let us formulate the special cases of maximal observables and maximal subalgebras
  as a corollary:
  \begin{cor}
    Let $X$ an observable maximal in $\alg{X}$, then $H(X)\geq H(\alg{X})$.
    Let $\alg{X}'$ a subalgebra maximal in $\alg{X}$, then $H(\alg{X}')\geq H(\alg{X})$. \qed
  \end{cor}
  An application of this is in the proof of
  \begin{satz}
    \label{satz:triangle}
    Let $\alg{X},\alg{Y}$ compatible, $\rho$ any state. Then
    $|H(\alg{X})-H(\alg{Y})|\leq H(\alg{XY})$.
  \end{satz}
  \proof Like in the previous theorem we may assume that $\rho$ is a state
  on $\alg{X}\otimes\alg{Y}$, and by symmetry we have to prove that
  $$H(\alg{X})-H(\alg{Y})\leq H(\alg{XY})$$
  If we think of $\alg{X}$ and $\alg{Y}$ as sums of full operator algebras,
  say $\alg{X}=\bigoplus_i \alg{L}({\cal H}_i)$, $\alg{Y}=\bigoplus_j \alg{L}({\cal K}_j)$,
  then embedding them into $\alg{L}(\bigoplus_i{\cal H}_i)$, $\alg{L}(\bigoplus_j{\cal K}_j)$,
  respectively, does not change the entropies involved (because the subalgebras are maximal).
  Thus we may assume that $\alg{X}=\alg{L}({\cal H})$, $\alg{Y}=\alg{L}({\cal K})$.
  Now consider a \emph{purification} $\ket{\psi}$ of $\rho$ on
  the Hilbert space ${\cal H}\otimes{\cal K}\otimes{\cal L}$ (see
  e.g.~\cite{schumacher:coherent}): this means
  $\rho=\tr_{\alg{L}({\cal L})}\ket{\psi}\bra{\psi}$.
  Now by theorem~\ref{satz:pure:common:state}
  $H(\alg{X})=H(\alg{YZ})$, $H(\alg{XY})=H(\alg{Z})$, and the assertion
  follows from subadditivity theorem~\ref{satz:strong:subadd}:
  $H(\alg{YZ})\leq H(\alg{Y})+H(\alg{Z})$.  \qed

  \subsection{Information}
  The following inequality for mutual information is a straightforward generalization
  of the Holevo bound~\cite{holevo:bound}, see also next section~\ref{sec:qmwc}:
  \begin{satz}
    \label{satz:bound}
    Let $X,Y$ be compatible observables with values in compatible $*$--subalgebras $\alg{X},\alg{Y}$,
    respectively. Then
    $${I}(X\wedge Y)\leq{I}(\alg{X}\wedge Y)\leq{I}(\alg{X}\wedge\alg{Y})$$
    (Conditions of equality!).
  \end{satz}
  \proof Consider the diagram
    \begin{equation*}\begin{CD}
         \alg{B}(\Omega_X)      @>{X}>>                  \alg{X}                 @=          \alg{X}\\
          @V{}VV                                          @V{}VV                          @V{\varphi}VV\\
    \alg{B}(\Omega_X)\otimes\alg{B}(\Omega_Y)
                                @>{X\otimes\id}>> \alg{X}\otimes\alg{B}(\Omega_Y)
                                                                        @>{\id\otimes Y}>>
                                                                                       \alg{X}\otimes\alg{Y}
                                                                                                   @>{\mu}>>\alg{A}\\
          @A{}AA                                          @A{}AA                          @A{\varphi'}AA\\
         \alg{B}(\Omega_Y)      @=                   \alg{B}(\Omega_Y)         @>{Y}>>        \alg{Y}
    \end{CD}\end{equation*}
    and apply the Lindblad--Uhlmann monotonicity
    theorem~\ref{sec:povm}.\ref{satz:monoton} twice, with
    $\mu_*(\rho)$ and the maps $(\id\otimes Y)_*$ and $(X\otimes\id)_*$, one after the other.\qed
  \par\medskip
  This can be greatly extended: for example if $\alg{X}\subset\alg{X}'$,
  $\alg{Y}\subset\alg{Y}'$, then
  $${I}(\alg{X}\wedge\alg{Y})\leq{I}(\alg{X}'\wedge\alg{Y}')$$
  The most general form is
  $${I}(\psi_1\circ\varphi_1\wedge\psi_2\circ\varphi_2)\leq{I}(\psi_1\wedge\psi_2)$$
  in the diagram
  \begin{equation*}\begin{CD}
          \alg{A}_1' @>{\varphi_1}>>                      \alg{A}_1            @>{\psi_1}>> \alg{A} \\
            @V{}VV                                         @V{}VV                             @| \\
    \alg{A}_1'\otimes\alg{A}_2' 
                     @>{\varphi_1\otimes\varphi_2}>> \alg{A}_1\otimes\alg{A}_2 @>{\psi=\psi_1\psi_2}>> \alg{A} \\
            @A{}AA                                         @A{}AA                             @| \\
          \alg{A}_2' @>{\varphi_2}>>                      \alg{A}_2            @>{\psi_2}>> \alg{A}
  \end{CD}\end{equation*}
  \begin{satz}
    \label{satz:info:subadd}
    Let $\alg{X}_1,\alg{X}_2,\alg{Y}_1,\alg{Y}_2$ compatible $*$--subalgebras of $\alg{A}$,
    $\rho$ a state on $\alg{A}$. Then
    $$I(\alg{X}_1\alg{X}_2\wedge\alg{Y}_1\alg{Y}_2)\leq
              I(\alg{X}_1\wedge\alg{Y}_1)+I(\alg{X}_2\wedge\alg{Y}_2)$$
    if $I(\alg{Y}_1\wedge\alg{X}_2\alg{Y}_2|\alg{X}_1)=0$ and
    $I(\alg{Y}_2\wedge\alg{X}_1\alg{Y}_1|\alg{X}_2)=0$ (i.e. $\alg{Y}_k$ is
    \emph{independent} from the other subalgebras conditional on $\alg{X}_k$).
  \end{satz}
  \proof
    First observe that the conditional independence mentioned,
    $I(\alg{Y}_1\wedge\alg{X}_2\alg{Y}_2|\alg{X}_1)=0$, is equivalent
    to $H(\alg{Y}_1|\alg{X}_1\alg{X}_2\alg{Y}_2)=H(\alg{Y}_1|\alg{X}_1)$.
    By theorem~\ref{satz:more:knowledge} we then have also
    $H(\alg{Y}_1|\alg{X}_1\alg{X}_2)=H(\alg{Y}_1|\alg{X}_1)$.
    Now observe (with the obvious chain rule)
    \begin{equation*}\begin{split}
      H(\alg{Y}_1\alg{Y}_2|\alg{X}_1\alg{X}_2)
                         &= H(\alg{Y}_1|\alg{X}_1\alg{X}_2\alg{Y}_2)+H(\alg{Y}_2|\alg{X}_1\alg{X}_2)\\
                         &= H(\alg{Y}_1|\alg{X}_1)+H(\alg{Y}_2|\alg{X}_2)
    \end{split}\end{equation*}
    and hence
    \begin{equation*}\begin{split}
      I(\alg{X}_1\alg{X}_2\wedge\alg{Y}_1\alg{Y}_2)
                         &=    H(\alg{Y}_1\alg{Y}_2)-H(\alg{Y}_1\alg{Y}_2|\alg{X}_1\alg{X}_2) \\
                         &\leq H(\alg{Y}_1)+H(\alg{Y}_2)-H(\alg{Y}_1|\alg{X}_1)-H(\alg{Y}_2|\alg{X}_2) \\
                         &=I(\alg{X}_1\wedge\alg{Y}_1)+I(\alg{X}_2\wedge\alg{Y}_2)
    \end{split}\end{equation*}
    where we have used the subadditivity of von Neumann entropy
    theorem~\ref{satz:strong:subadd}.
  \qed
  \\
  The same obviously applies if we have $n$ $*$--subalgebras $\alg{X}_k$, and
  $n$ $\alg{Y}_k$, all compatible, and if $\alg{Y}_k$ is independent from the others
  give $\alg{X}_k$, i.e. for all $k$
  $$H(\alg{Y}_k|\alg{X}_1\cdots\alg{X}_n\alg{Y}_1\cdots\widehat{\alg{Y}_k}\cdots\alg{Y}_n)
         =H(\alg{Y}_k|\alg{X}_k)$$
  \begin{cor}
    \label{cor:info:subadd}
    Let $\alg{X}_1,\ldots,\alg{X}_n$, $\alg{Y}_1,\ldots,\alg{Y}_n$ C${}^*$--algebras,
    $\alg{X}_i=\C\fset{X}_i$, 
    $\alg{A}=\alg{X}_1\otimes\cdots\otimes\alg{X}_n\otimes\alg{Y}_1\otimes\cdots\otimes\alg{Y}_n$.
    and a probability distribution $P$ on $\fset{X}_1\times\cdots\times\fset{X}_n$.
    Then with the state
    $$\gamma=\sum_{x_i\in\fset{X}_i} P(x_1,\ldots,x_n)x_1\otimes\cdots\otimes x_n\otimes
                                       W_{x_1}\otimes\cdots\otimes W_{x_n}$$
    on $\alg{A}$ (where $P$ is a probability on $\fset{X}_1\times\cdots\times\fset{X}_n$ and
    $W$ maps the $\fset{X}_i$ to states on $\alg{Y}_i$):
    $$I(\alg{X}_1\cdots\alg{X}_n\wedge\alg{Y}_1\cdots\alg{Y}_n)\leq\sum_k I(\alg{X}_k\wedge\alg{Y}_k)$$
  \end{cor}
  \proof We only have to check the conditional independence, which is left to the reader.
  \qed
  \par\medskip
  We note another simple estimate for the mutual information:
  \begin{satz}
    \label{satz:info:upperbound}
    For compatible $*$--subalgebras $\alg{X},\alg{Y}$:
    $I(\alg{X}\wedge\alg{Y})\leq 2\min\{H(\alg{X}),H(\alg{Y})\}$
  \end{satz}
  \proof Put together the formula
  $I(\alg{X}\wedge\alg{Y})=H(\alg{X})-H(\alg{X}|\alg{Y})$ and the simple
  estimate $H(\alg{X}|\alg{Y})\geq -H(\alg{X})$ from theorem~\ref{satz:triangle}.
  \qed

  \subsection{Conditional entropy}
  \begin{satz}
    \label{satz:conditional:entropy}
    Let $\varphi:\alg{X}\rightarrow\alg{A}$, $\psi:\alg{Y}\rightarrow\alg{A}$ compatible
    quantum operations
    with $\alg{X}$ or $\alg{Y}$ commutative. Then $H(\varphi|\psi)\geq 0$.
  \end{satz}
  \proof
    Let $\sigma=(\varphi\psi)_*\rho$, then by definition and
    lemma~\ref{sec:subalgebra}.\ref{lemma:onto:hom}
    $$H(\varphi|\psi)=H(\sigma)-H(\tr_\alg{Y}\sigma)$$
    \emph{First case}: $\alg{X}$ is commutative, so we can write
    $\sigma=\sum_x Q(x)[x]\otimes\tau_x$ with a distribution $Q$ on $\fset{X}$, and
    states $\tau_x$ on $\alg{Y}$. Obviously
    $H(\sigma)=H(Q)+\sum_x Q(x)H(\tau_x)$, and
    $\tr_\alg{Y}\sigma=\sum_x Q(x)[x]=Q$, and hence
    $H(\varphi|\psi)=\sum_x Q(x)H(\tau_x)\geq 0$.\\
    \emph{Second case}: $\alg{Y}$ is commutative, so we can write
    $\sigma=\sum_x Q(x)[x]\tau_x\otimes[x]$, like in the first case. $H(\sigma)$
    is calculated as before, but now
    $\tr_\alg{Y}\sigma=\sum_x Q(x)\tau_x=Q\tau$, and
    \begin{equation*}\begin{split}
      H(\varphi|\psi) &= H(Q)-(H(Q\tau)-\sum_x Q(x)H(\tau_x)) \\
                      &= H(Q)-I(Q,\tau)\geq 0
    \end{split}\end{equation*}
    (see section~\ref{sec:qmwc}, for the last line theorem~\ref{sec:qmwc}.\ref{satz:holevo}).
  \qed
  \begin{bem}
    From the proof we see that the commutativity of $\alg{X}$ or
    $\alg{Y}$ enters in the representation
    of $\sigma$ as a particular separable state with respect to the subalgebras $\alg{X}$,
    $\alg{Y}$ (see definition below),
    namely with one party admitting common diagonalization of her states.
    We formulate as a conjecture the more general:\\
    $H(\alg{X}|\alg{Y})\geq 0$ if
    $\rho$ is separable with respect to $\alg{X}$ and $\alg{Y}$.\\
    From this it would follow that in this case
    $I(\alg{X}\wedge\alg{Y})\leq\min\{H(\alg{X}),H(\alg{Y})\}$
    (see theorem~\ref{satz:info:upperbound}), which we now only get from the
    commutativity assumption.
  \end{bem}
  \begin{defi}
    \label{defi:separable}
    Call $\rho$ \emph{separable with respect to} the compatible $*$--subalgebras
    $\alg{X}_1,\ldots,\alg{X}_m$ of $\alg{A}$, if, for the natural multiplication map
    $\mu:\alg{X}_1\otimes\cdots\otimes\alg{X}_m\rightarrow\alg{A}$, $\mu_*\rho$
    is a separable state on $\alg{X}_1\otimes\cdots\otimes\alg{X}_m$, i.e. a convex
    combination of product states $\sigma_1\otimes\cdots\otimes\sigma_m$,
    $\sigma_i\in\alg{S}(\alg{X}_i)$. If $\mu_*\rho$ is a product state, we call
    also $\rho$ a \emph{product state with respect to} $\alg{X}_1,\ldots,\alg{X}_m$.
  \end{defi}
  \begin{satz}[Knowledge decreases uncertainty]
    \label{satz:more:knowledge}
    Let $\varphi:\alg{X}\rightarrow\alg{A}$, $\psi:\alg{Y}\rightarrow\alg{A}$ compatible
    quantum operations, and $\varphi':\alg{X}'\rightarrow\alg{X}$ any quantum operation.
    Then $H(\psi|\varphi)\leq H(\psi|\varphi\circ\varphi')$, in particular
    $H(\psi|\varphi)\leq H(\psi)$.
  \end{satz}
  \proof
    The inequality is obviously equivalent to
    $I(\psi\wedge\varphi)\geq I(\psi\wedge\varphi\circ\varphi')$, i.e. theorem~\ref{satz:bound}.
  \qed
  \\ \medskip
  Defining $h(x)=-x\log x-(1-x)\log(1-x)$ for $x\in[0,1]$ we have the famous
  \begin{satz}[Fano inequality]
    \label{satz:fano:inequality}
    Let $\rho$ a state on $\alg{A}$, and
    $\alg{Y}$ be a $*$--subalgebra of $\alg{A}$, compatible with the observable $X$ (indexed by
    $\fset{X}$). Then
    for any observable $Y$ with values in $\alg{Y}$ the probability that ``$X\neq Y$'',
    i.e. $P_e=1-\sum_j \tr(\rho X_jY_j)$, satisfies
    $$H(X|\alg{Y})\leq h(P_e)+P_e\log(|\fset{X}|-1)$$
  \end{satz}
  \proof
    By the previous theorem~\ref{satz:more:knowledge} it suffices to prove the
    inequality with $H(X|Y)$ instead of $H(X|\alg{Y})$. But then we have the classical
    Fano inequality: the uncertainty on $X$ given $Y$ may be estimated by the
    uncertainty of the event that they are equal plus the uncertainty on the
    value of $X$ if they are not.
  \qed
  \begin{cor}
    \label{cor:fano:inequality}
    Let $\alg{X}$ a commutative $*$--subalgebra compatible with $\alg{Y}$, and $X$
    the (uniquely determined) maximal observable on $\alg{X}$, $P_e$ as in the
    theorem, then
    $$H(\alg{X}|\alg{Y})\leq h(P_e)+P_e\log(\tr\supp(\rho|_{\alg{X}})-1)$$
  \end{cor}
  \proof
    First observe that $H(\alg{X}|\alg{Y})=H(X|\alg{Y})$. To apply the theorem
    we only have to restrict the range of $X$ to those values that are actually
    assumed.
  \qed

\section{Quantum channels}
  \label{sec:qmwc}
  \setcounter{defi}{0}
  
  \subsection{General remarks}
  \label{subsec:general}
  We consider in the following only quantum channels with a priori fixed input states
  (i.e. \emph{classical--quantum} channels after Holevo~\cite{holevo:channels}).
  Formally such a system may be described by the collection $(W_x|x\in{\cal X})$
  of states with $W_x$ appearing at the receiving end when $x$ is sent. This may
  also be described by its linear extension
  $W:{\C{\cal X}}_*\rightarrow\alg{Y}_*$, a trace preserving quantum operation
  (this is the only occasion where we omit the subscript
  $*$ for a quantum map between state spaces).\\
  Side remark: the most general quantum channel appears if we allow at the left any
  C${}^*$--algebra instead of the commutative one. In this case we are free to choose
  input states --- in  general from a continuum. Even more, we may (in block coding)
  use entangled states. For simplicity, and because of some unsolved problems in the
  more general case we decided here to stay with classical--quantum channels.\\
  This idea of a channel as a process, after choosing a distribution $P$ on
  ${\cal X}$ (i.e. a state on $\alg{X}=\C{\cal X}$), which is an average input,
  leads to the notions of the average output $PW=W(P)=\sum_{x\in{\cal X}} P(x)W_x$
  and the mutual information $I(P,W)=H(PW)-\sum_{x\in{\cal X}} P(x)H(W_x)$.\\
  Whereas this is a physically perfectly reasonable model with its appropriate
  ideas, looking at classical information theory we see that there is also another
  way of thinking about channels: namely as stochastic two--end systems, one end
  of which is declared the sender, the other the receiver (even though formally the
  thing is symmetric), and their respective input and output distributions
  are marginals of some joint distribution (which reflects the dependence of the
  output on the input). To model this with quantum systems
  define the \emph{channel state} $\gamma=\sum_x P(x) x\otimes W_x$ on
  $\alg{X}\otimes\alg{Y}$. Notice that we (abstractly, and somewhat unnaturally) divided
  the system into two: its past and its future, and $\gamma$ describes the
  correlation between them. Obviously $P$ and $PW$ are obtained as marginals,
  by tracing over $\alg{Y}$, $\alg{X}$, respectively. In fact it is an easy exercise
  to verify that $I(P,W)=I(\alg{X}\wedge\alg{Y})$.\\
  This second point of view (and its connection to the first, which was
  noticed before by Hall~\cite{hall:context}
  in his investigation of what he calls \emph{context mappings})
  was the motive for the whole presentation in the preceding sections: to phrase the
  information and entropy concepts initially defined in the context of processing
  states via quantum operations in a ``static'' model that allows for the use of
  observables (i.e. random variables), and comparison with certain subalgebras.
  
  \subsection{Multiway channels}
  \label{subsec:muliway}
  In the sequel we will also consider a more general channel: we call it the
  \emph{(all--to--all) quantum multiway channel} with $s$ senders and $r$ receivers
  (or the $r$--\emph{fold compound multiple access channel}), and
  it consists of $s$ commutative C${}^*$--algebras $\alg{X}_1,\ldots,\alg{X}_s$
  (say $\alg{X}_i=\C\fset{X}_i$, and let $\alg{X}=\alg{X}_1\otimes\cdots\otimes\alg{X}_s$),
  a quantum operation $W:\alg{X}\rightarrow\alg{Y}$, and $r$ compatible
  $*$--subalgebras $\alg{Y}_1,\ldots,\alg{Y}_r$ of $\alg{Y}$. The idea
  here is that the $\alg{X}_i$ are the senders, the $\alg{Y}_j$ the receivers,
  and each sender wants to send the same message to every receiver, with
  small error probability. This we formalize in the notion of an $(n,\bar{\epsilon})$--code
  which consists of $s$ mappings $f_i:\fset{M}_i\rightarrow\fset{X}_i^n$
  with finite set $\fset{M}_i$, and $r$ decoding observables $Y_j$, indexed by
  $\fset{M}_1'\times\cdots\times\fset{M}_s'\supset\fset{M}_1\times\cdots\times\fset{M}_s$,
  with values in $\alg{Y}_j^{\otimes n}$ (and so these are automatically 
  compatible) with the $r$ (average) error probabilities
  $$\bar{e}_j(f_1,\ldots,f_s,Y_j)=1-\frac{1}{|\fset{M}_1|\cdots|\fset{M}_s|}
      \sum_{\forall i:m_i\in\fset{M}_i}\tr\left({W^{\otimes n}(f(m_1),\ldots,f(m_s))
             Y_{j,m_1\ldots m_s}}\right)$$
  all being at most $\bar{\epsilon}$. The \emph{rate} of the code is the tuple $(R_1,\ldots,R_s)$
  with $R_i=\dfrac{1}{n}\log|\fset{M}_i|$. The problem is then to determine the
  \emph{capacity regions} ${\bf R}(\bar{\epsilon})$, i.e. the set of all achievable $s$--tuples with error
  probability $\bar{\epsilon}$ (where \emph{achievable} means that for infinitely many
  $n$ there exist $(n,\bar{\epsilon})$--codes with rate tuples converging to the given tuple),
  or more realistically ${\bf R}=\bigcap_{\bar{\epsilon}>0} {\bf R}(\bar{\epsilon})$
  (which is usually called \emph{the} capacity region).
  Obviously this consists of two parts: first to exhibit the
  existence of codes with certain rate, second bounds on the rate for any code.\\
  A little history: with classical communication the multiway channel was first considered
  by Shannon~\cite{shannon:multiway}, and the exact determination of the capacity region
  was done by Ahlswede~\cite{ahlswede:mac,ahlswede:2s2r}. There are of course even more
  general multi--user communication models, most of which are unsolved: a good overview
  is in the paper~\cite{elgamal:cover} by El~Gamal and Cover. Quantum channels for single
  sender and receiver were all around since the sixties, but the first formal definitions
  seem to have been given by Holevo~\cite{holevo:bound,holevo:channels}. The quantum
  multiway channel as defined here is a slightly smoothed presentation of the definition
  by Allahverdyan and Saakian~\cite{allahverdyan:saakian} (where the channel is a general
  quantum map).\\
  Before we can tackle this problem (of which we will solve in this paper only the second
  part, giving bounds) we have to collect a few facts.\\
  The following is a corollary to the information inequality:
  \begin{satz}[Holevo bound]
    \label{satz:holevo}
    For any classical--quantum channel $W:\fset{X}\rightarrow\alg{S}(\alg{Y})$,
    any distribution $P$ on ${\cal X}$, and any observable $Y$ on $\alg{Y}$
    $$I(P,W)\geq I(P,Y_*\circ W)$$
    More generally, for any completely positive quantum operation
    $\varphi:\alg{Z}\rightarrow\alg{Y}$ one has $I(P,W)\geq I(P,\varphi_*\circ W)$.
    In particular $I(P,W)\leq I(P,\id)=H(P)$.
  \end{satz}
  \proof All ingredients are already known: we define a channel state
  $\gamma=\sum_x P(x)[x]\otimes [x]$ on $\C\fset{X}\otimes\C\fset{X}$ and observe
  that $I(P,\id)=I(\id_1\wedge\id_2)=H(P)$. Now to apply the information bound
  let $\psi:\alg{Y}\rightarrow\C\fset{X}$ such that $W=\psi_*$:
  $$
  \begin{array}{ccccc}
    I(\id_1\wedge\id_2) & \geq & I(\id_1\wedge\psi) & \geq & I(\id_1\wedge\psi\circ\varphi) \\
          \|            &      &        \|          &      &             \|                 \\
         H(P)           &      &      I(P,W)        &      &    I(P,\varphi_*\circ W)
  \end{array}
  $$
  \qed
  \par\medskip
  The formulation of the Holevo bound is of course in the manner of a data
  processing inequality, data processing in the sense of composition of two
  quantum operations. We can also formulate it in the language of observables,
  just like for classical correlated random variables:\\
  For this consider the following state on $\alg{X}\otimes\alg{Y}\otimes\alg{Z}$
  $$\gamma=\sum_{x\in{\cal X}} P(x)x\otimes W_x\otimes\varphi_*(W_x)$$
  which represents the correlation of the three stages of the system:
  preparation, reception, and detection of the signal (again note that this is
  artificial in the material sense). The data processing inequality now
  is in the familiar form $I(\alg{X}\wedge\alg{Z})\leq I(\alg{X}\wedge\alg{Y})$.
  For proof check identity of the information terms with those in the
  Holevo bound.\\
  We might want to try to imitate the well known classical proof
  for random variables: by obvious chain rules
  \begin{equation*}\begin{split}
    I(\alg{X}\wedge\alg{Y}\alg{Z}) &= I(\alg{X}\wedge\alg{Y}|\alg{Z})+I(\alg{X}\wedge\alg{Z})\\
                                   &= I(\alg{X}\wedge\alg{Z}|\alg{Y})+I(\alg{X}\wedge\alg{Y})
  \end{split}\end{equation*}
  Since $I(\alg{X}\wedge\alg{Y}|\alg{Z})\geq 0$ the inequality will be proved if we
  could show that $I(\alg{X}\wedge\alg{Z}|\alg{Y})=0$: but this is not true, as we will
  show immediately by example! Before we do that however let us discuss our definition
  of $\gamma$. Observe that it not even in the classical case reflects the dependence
  of $\alg{Z}$ on $\alg{Y}$ correctly: $W_x$ is a sum of pure (deterministic) states,
  say $W_x=\sum_y W(y|x) V_y$ (classically of course this is unique, and
  $V_y$ \emph{is} just $y$), and $\varphi_*$ invidually transforms these states. Thus a
  better choice whould be
  $$\gamma=\sum_{x,y} P(x)W(y|x)x\otimes V_y\otimes \varphi_*(V_y)$$
  Note that this does not change $I(\alg{X}\wedge\alg{Z})$ or $I(\alg{X}\wedge\alg{Y})$.
  On the other hand the decomposition of $W_x$ is no longer unique in the
  quantum case. In our example however the $W_x$ will be pure, so there is
  in fact no question of decomposition:
  \begin{expl}
    {\rm
    Consider a binary channel, i.e. $\fset{X}=\{0,1\}$, $\alg{Y}=\alg{L}(\C^2)$. In $\C^2$
    fix an orthonormal basis $\ket{0},\ket{1}$ and let $\ket{+}=\dfrac{1}{\sqrt{2}}(\ket{0}+\ket{1})$.
    Let $W_0=\ket{0}\bra{0}$, $W_1=\ket{+}\bra{+}$, and $P$ the uniform distribution.\\
    In the first scenario let $\varphi_*=\id$, so
    $$\gamma=\frac{1}{2}[0]\otimes\ket{0}\bra{0}\otimes\ket{0}\bra{0}
            +\frac{1}{2}[1]\otimes\ket{+}\bra{+}\otimes\ket{+}\bra{+}$$
    and a short calculation shows
    $$\begin{array}{ll}
      H(\alg{X}|\alg{Y})        &=1-h(\cos^2\frac{\pi}{8})\approx .399\\
      H(\alg{X}|\alg{Y}\alg{Z}) &=1-h(\cos^2\frac{\pi}{6})\approx .189
      \end{array}$$
    The difference
    is easily explained: in the second quantity one has access to two clones of the
    original state $W_x$, so identifying $x$ is better possible.\\
    This principle of doubly using quantum information in a forbidden way
    still is possible even if we insist that $\varphi$ should be a measurement:
    in the second scenario $\varphi_*$ is the external operation of a von
    Neumann measurement in basis
    $$\ket{u}=\cos\frac{\pi}{8}\ket{0}-\sin\frac{\pi}{8}\ket{1},\quad
      \ket{v}=\sin\frac{\pi}{8}\ket{0}+\cos\frac{\pi}{8}\ket{1}$$
    Thus (with $\alpha=\cos^2\dfrac{\pi}{8}=\dfrac{1+\sqrt{1/2}}{2}$)
    $$\gamma=\frac{1}{2}[0]\otimes W_0\otimes(\alpha    [0]+(1-\alpha)[1])+
             \frac{1}{2}[1]\otimes W_1\otimes((1-\alpha)[0]+\alpha    [1])$$
    and an easy calculation shows (with
    $\beta=\dfrac{1+\sqrt{1-2\alpha(1-\alpha)}}{2}=\dfrac{1+\sqrt{3/4}}{2}$)
    $$\begin{array}{ll}
      H(\alg{X}|\alg{Y})        &=1-h(\alpha)\approx .399\\
      H(\alg{X}|\alg{Y}\alg{Z}) &=1-h(\beta) \approx .246
      \end{array}$$
    Again the reason for the failure is the same (which is unknown in the classical theory):
    in $\gamma$ we consider states as coexistent which never can coexist,
    because the third stage evolves from the second by an operation (a measurement) which must
    needs disturb the system: we neglected this very fact in constructing $\gamma$, and we had to:
    otherwise we could not have incorporated both stages of the evolution, the one after
    $W$, and the one after $\varphi_*$.}
  \end{expl}
  After this digression we turn to an application of the Holevo bound: with the above notation
  \begin{satz}[Upper capacity bounds]
    \label{satz:s2r:upperbound}
    The capacity region of the quantum multiway channel is contained in the
    closure of all nonnegative
    $(R_1,\ldots,R_s)$ satisfying
    $$\forall J\subset[s],j\in[r]\qquad \sum_{i\in J} R_i
            \leq\sum_u q_u I_{\gamma_u}\left(\alg{X}(J)\wedge\alg{Y}_j|\alg{X}(J^c)\right)$$
    for some channel states $\gamma_u$ (belonging to appropriate input distributions) and 
    $q_u\geq 0$, $\sum_u q_u=1$.
  \end{satz}
  \proof Assume an $(n,\bar{\epsilon})$--code $(f_1,\ldots,f_s,Y_1,\ldots,Y_r)$.
    Then the uniform distribution on the codewords induces a channel state
    $\gamma$ on $(\alg{X}_1\cdots\alg{X}_s\alg{Y})^{\otimes n}$. Its restriction
    to the $u$--th copy in this tensor power will be denoted $\gamma_u$.
    Let $j\in[r]$, $J\subset[s]$.
    By Fano inequality~\ref{sec:bound}.\ref{satz:fano:inequality} (and corollary) we have
    $$H(\alg{X}^{\otimes n}(J)|\alg{Y}^{\otimes n}_j\alg{X}^{\otimes n}(J^c))
                                                           \leq 1+\bar{\epsilon}\cdot nR(J)$$
    With
    \begin{equation*}\begin{split}
      H(\alg{X}^{\otimes n}(J)|\alg{Y}^{\otimes n}_j\alg{X}^{\otimes n}(J^c))
                      &=H(\alg{X}^{\otimes n}(J))
                           -I(\alg{X}^{\otimes n}(J)\wedge\alg{Y}^{\otimes n}_j\alg{X}^{\otimes n}(J^c))\\
                      &=nR(J)-I(\alg{X}^{\otimes n}(J)\wedge\alg{Y}^{\otimes n}_j\alg{X}^{\otimes n}(J^c))
    \end{split}\end{equation*}
    we conclude (with theorem~\ref{sec:bound}.\ref{satz:info:subadd} and corollary)
    \begin{equation*}\begin{split}
      (1-\bar{\epsilon})R(J) &\leq \frac{1}{n}+\frac{1}{n}
                            I_\gamma(\alg{X}^{\otimes n}(J)\wedge\alg{Y}^{\otimes n}_j\alg{X}^{\otimes n}(J^c))\\
                             &\leq \frac{1}{n}+\frac{1}{n}\sum_{u=1}^{n}
                                                  I_{\gamma_u}(\alg{X}(J)\wedge\alg{Y}_j\alg{X}(J^c))
    \end{split}\end{equation*}
  \qed
  \begin{bem}
    \label{bem:ahlswede:cap}
    In the case of classical channels the region described in the theorem is the exact capacity
    region (i.e. all the rates there are achievable), as was first proved by
    Ahlswede~\cite{ahlswede:mac,ahlswede:2s2r}.
  \end{bem}
  \begin{bem}
    The significance of the Holevo bound lies in that we can with it
    and the Fano inequality derive an upper bound on the capacity of a quantum
    channel.
    Holevo~\cite{holevo:capacity} and independently Schumacher and
    Westmoreland~\cite{schumacher:westmoreland} recently showed that in the case
    $\alg{Y}=\alg{L}({\cal H})$ this bound can be achieved.
    In~\cite{winter:qmac} achievability in the 
    case of the multiple access channel ($r=1$) and for general $\alg{Y}$ is demonstrated.\\
    We conjecture that also in the general case of $r>1$ 
    the theorem gives already the right capacity region.
  \end{bem}

  \subsection{Broadcast channels}
  \label{subsec:broadcast}
  To end this section let us think a bit about the quantum analog of the
  \emph{broadcast channel} (see also recent work by Allahverdyan and
  Saakian~\cite{allahverdyan:saakian:qbroad}):
  suppose a sender wants to transmit messages from two sets to two receivers --- over the
  same quantum channel (like a TV--station with several programs). Receiver $1$ is interested in
  part $1$ of the message, receiver $2$ in part $2$, both in a common part $0$.
  A model of this situation is
  a map $W:\fset{X}\rightarrow\alg{S}(\alg{Y})$ for the channel, two $*$--subalgebras
  $\alg{Y}_1,\alg{Y}_2$ of $\alg{Y}$ for the two receivers: the triple
  $(W,\alg{Y}_1,\alg{Y}_2)$ we call a broadcast channel.
  If these subalgebras are compatible we call the system \emph{plug--and--play}
  (because then each receiver may choose any observable without interfering
  with the other. In the other case they may have to agree on compatible observables, or
  prescribe the order of access to the data).
  An $n$--block code for this channel is a triple $(f,D_1,D_2)$
  with a map $f:\fset{M}_0\times\fset{M}_1\times\fset{M}_2\rightarrow\fset{X}^n$ and compatible
  observables $D_i$ in $\alg{Y}_i$, indexed by
  $\fset{M}_0'\times\fset{M}_i'\supset\fset{M}_0\times\fset{M}_i$ ($i=1,2$).
  The (maximum) error probability of the code is
  $$e(f,D_1,D_2)=\max\{1-\tr(W_{f(m_0,m_1,m_2)}D_{1,m_0m_1}D_{2,m_0m_2})|
                                                  m_i\in\fset{M}_i,\ i=0,1,2\}$$
  (and analogously the average error probability $\bar{e}$). If it is at most $\epsilon$
  we speak of a $(n,\epsilon)$--code ($(n,\bar{\epsilon})$--code, respectively).
  The capacity of the code is, as expected, the triple
  $(R_1,R_0,R_2)=(\dfrac{1}{n}\log|\fset{M}_1|,\dfrac{1}{n}\log|\fset{M}_0|,
             \dfrac{1}{n}\log|\fset{M}_2|)$, and the problem
  is to determine the capacity region. This is a problem exceedingly difficult,
  not even solved completely in the classical case ($\alg{Y}$ commutative).
  \\
  Thus we may
  consider a restricted situation, which has in the classical case a complete
  solution: the \emph{degraded} broadcast channel: here the line to receiver $2$
  ``factors'' through $1$, i.e. the degraded broadcast channel is a triple
  $(W,\varphi_*,\alg{Y}_1)$ with $W$ as above, and a quantum operation
  $\varphi_*:\alg{Y}_*\rightarrow\alg{Y}_{2*}$.
  Receiver $1$ is the $*$--subalgebra $\alg{Y}_1$
  of $\alg{Y}$, receiver $2$ the algebra $\alg{Y}_2$.
  This links with the previous explanation via the definitions
  $W_2=\varphi_*\circ W$, $W_1=\imath_*\circ W$ (for the inclusion
  $\imath:\alg{Y}_1\hookrightarrow\alg{Y}$). This however gives not the correct
  picture because this model is manifestly not plug--and--play:
  The second receiver has to take what the first left to him.
  Formally: an $n$--block code
  $(f,\tilde{D}_1,D_2)$ now consists
  of $f$ as before, and also the observable $D_2$ on $\alg{Y}_2^{\otimes n}$, and a subtle modification
  of $D_1$ to the operation $\tilde{D}_1=\psi_*$ for a quantum operation
  \begin{equation*}\begin{array}{rll}
    \psi:\C\fset{M}_0'\otimes\C\fset{M}_1'\otimes\alg{Y}^{\otimes n} & \rightarrow & \alg{Y}^{\otimes n}\\
                                    m_0\otimes m_1\otimes A & \mapsto     & E_{m_0m_1}^* A E_{m_0m_1}
  \end{array}\end{equation*}
  with $E_{m_0m_1}\in\alg{Y}_1^{\otimes n}$.
  Obviously $\tr_{\alg{Y}^{\otimes n}}\circ\tilde{D}_1=D_1$
  for the observable $D_1$ indexed by $\fset{M}_0'\times\fset{M}_1'$ and consisting of the
  operators $D_{1,m_0m_1}=E_{m_0m_1}E_{m_0m_1}^*$. With this we can formulate the
  error probability:
  $$e(f,\tilde{D}_1,D_2)=\max\{1-\tr(E_{m_0m_1}^*W_{f(m_0,m_1,m_2)}E_{m_0m_1}D_{2,m_0m_2})|
                                                 m_i\in\fset{M}_i,\ i=0,1,2\}$$
  (analogous for the average error probability).
  In direct analogy with the classical situation we present the following
  \begin{conj}
    For $\alg{Y}_1=\alg{Y}$ the rate region is the convex hull of the triples
    $(R_1,R_0,R_2)$ with
    $$\begin{array}{rl}
        R_1          &\leq I(V(\cdot|u),W|Q) \\
        R_0+R_2      &\leq I(Q,\varphi_*\circ W\circ V)\\
        R_1+R_0+R_2  &\leq I(QV,W)
      \end{array}$$
    where $Q$ is a distribution on a finite set $\fset{U}$, and $V$ a classical
    channel from $\fset{U}$ to $\fset{X}$.
  \end{conj}

  \subsection{Open problems}
  \label{subsec:open}
  \begin{bem}
    Meaning of theorem~\ref{sec:bound}.\ref{satz:entropy:inequ}
    for coding theorems: The reason why for truely quantum channels
    one has strict inequality is that we cannot detect the $W_x$ optimally
    in one common basis (for simplicity assume that we only employ von Neumann
    measurements). Assume we chose an eigenbasis of $PW$, then we ``see''
    correctly the entropy $H(PW)$ of the output state, but for the letter
    states we introduce some additional entropy to their $H(W_x)$.
    Thus we get to low a mutual information because our measurements introduce
    noise. We want this noise increase
    to be small by choosing codewords appropriately, and then
    ``approximating'' with a von Neumann observable,
    all the codeword states nearly commute with. The problem here is to do
    this such that the von Neumann mutual information remains the same.\\
    Note that this is a different
    approach to coding than those used so far: there we directly construct codes
    approaching certain rate, using general observables. Here we would have
    a von Neumann observable approaching the Holevo bound, i.e. a classical
    channel for which we may construct codes by the known classical techniques.
  \end{bem}
  \begin{bem}
    For classical--quantum channels there does not appear to exist a reasonable
    notion of \emph{transpose} channel. If however we see a channel as a quantum map
    from any one system to another, then given an input state one can define
    formally a transpose channel under certain circumstances, see~\cite{ohya:petz}.
    This goes the opposite direction as the original channel, so in
    our case we get a measurement operation. It is to be explored whether
    this notion gives us new insight in the communication problem. In particular
    we may relate the classical--quantum channels with quantum--classical
    channels (i.e. fixed measurements, or if variable only product
    measurements). Maybe we can even prove that coding
    classical information with entangled states in quantum--quantum channels
    yields higher capacities...
  \end{bem}

\acknowledgements
  Thanks to Peter L\"ober for discussions during the course of this work, especially for
  pointing out to me the importance of strong subadditivity. Thanks also to MJW Hall for
  drawing my attention to his work.

\end{document}